\documentclass[pra,aps,unsortedaddress,superscriptaddress,amsmath,amssymb,twocolumn,floatfix]{revtex4}
    \usepackage{graphics}
    \usepackage{epstopdf}
   	\usepackage{epsfig}
	\usepackage{amsmath}
	\usepackage{makeidx}
	\usepackage{amsfonts}
	\usepackage[ansinew]{inputenc}
	\usepackage[usenames,dvipsnames]{pstricks}
	\usepackage{subfigure}
	\usepackage{pst-grad} 
	\usepackage{pst-plot} 
	\usepackage[colorlinks,hyperindex]{hyperref}
	\usepackage{xcolor}
	\hypersetup	
	{
		colorlinks,%
		citecolor=black,%
		linkcolor=black,%
		urlcolor=black,%
	}

\makeindex

\begin{document}

\title{Higher order non-linear effects in a Josephson parametric amplifier}

\author{Bogdan A. Kochetov}
\email{k.bogdan@uq.edu.au}
\author{Arkady Fedorov}
\affiliation{ARC Centre of Excellence for Engineered Quantum Systems, \\ The University of Queensland, St. Lucia QLD 4072, Australia}


\begin{abstract}
Non-linearity of the current-phase relationship of a Josephson junction is the key resource for a Josephson parametric amplifier (JPA), the only device in which the quantum limit has so far been achieved at microwave frequencies. A standard  approach to describe JPA takes into account only the lowest order (cubic) non-linearity resulting in a Duffing-like oscillator equation of motion or in a Kerr-type non-linearity term in the Hamiltonian. In this paper we derive the quantum expression for the gain of JPA including all orders of the Josephson junction non-linearity in the linear response regime. We then analyse gain saturation effect for stronger signals within semi-classical approach. Our results reveal non-linear effects of higher orders and their implications for operation of a JPA.


\end{abstract}


\maketitle

\section{Introduction}
\label{ch_I}
Non-linear nature of a Josephson junction has been recognized as a valuable resource for parametric amplification over several decades \cite{Likharev}. The earliest experiments with a microwave oscillator incorporating a Josephson junction as a non-linear inductance have incontrovertibly shown that these devices, called Josephson parametric amplifiers can operate as amplifiers with extremely low noise temperatures \cite{Yurke2} - \cite{Movshovich}. Rapid development of low-temperature experiments with superconducting qubits requesting the quantum limited amplifiers to control and measure a broad range of parameters of quantum systems and absence of suitable commercial amplifiers stimulated great interest in JPA in the recent decade. Recent experiments confirmed that JPA can work as an amplifier near quantum limit \cite{Castellanos} and can be used for a wide range of applications. These include measurement of a displacement of a nano-mechanical oscillator~\cite{Regal}, \cite{Teufel}, implementation of the high-fidelity single-shot measurement~\cite{Vijay1}-\cite{Lin}, quantum feedback~\cite{Vijay}, \cite{Riste} and feedforward~\cite{Steffen} for superconducting qubits, as well as generation of the two-mode squeezed states of  microwave photons \cite{Eichler2}.

A standard approach of a description of JPA takes into account only the leading order non-linearity originated from a Josephson junction. In particular, keeping the first non-linear term only in the Josephson current-phase relationship one gets an oscillator equation of motion with the cubic non-linearity analogous to the Duffing equation. In quantum approach the Hamiltonian acquires the Kerr-type non-linearity term coming from truncation of the energy of the Josephson junction. This simplification allows for an analytical solution for the steady state of an oscillator under strong pumping conditions. By linearising equation of motion for a weak external signal in the presence of a strong pump field one can obtain the analytical expressions for the gain of JPA (see for example \cite{Eichler1}). These formulas  provide reasonable agreement with experimental data but some discrepancies has been found, especially for low Q-factor oscillators \cite{Eichler1}. In addition, the linear response theory did not account for gain saturation effects in the presence of strong signals~\cite{Eichler1}, \cite{Rhoads} which are essential for the understanding of limitations on the dynamic range of an amplifier, a property of large practical importance. 
 
The main goal of our paper is to reveal and  analyse the non-linear effects in the JPA due to unavoidable presence of the higher non-linear terms in the current-phase relationship of a Josephson junction. First, we account for all non-linear terms in the steady state solution of a driven oscillator at the pump frequency. This allows us to derive the exact quantum expressions for the gain of a parametric amplifier in the linear amplification regime. Secondly, we derive and solve the equation of motions beyond the linear response theory  to account for gain saturation effects under semi-classical approximation taking into account all orders of non-linearity. After revealing non-linear effects of higher orders we discuss applicability of the obtained solutions and their implications for operation of JPA.

The paper is organized as follows. In Sec.~\ref{ch_E_H} we write down the Hamiltonian of a JPA and formulate the input-output theory to derive the exact equation of motion. The response for a strong classical pump field accounting for a complete $sine$ non-linearity is found in Sec.~\ref{ch_C_R}. In Sec.~\ref{ch_T_L} we derive equation for a weak quantum signal and solve it analytically in the frequency domain. Having found the JPA gain expression we study higher order non-linear effects. In Sec.~\ref{ch_G_S} we derive equation for an arbitrary strength quantum signal including both the fundamental non-linear terms and all non-linear terms and solve the equation numerically using semiclassical approach. The solution was used to describe the gain saturation effect in a JPA and estimate its dynamical range. Conclusions and final remarks are made in Sec.~\ref{ch_C}.

\section{Exact Hamiltonian and input-output relations}
\label{ch_E_H}
\begin{figure}[htbp]
\begin{center}
\includegraphics[width=0.45\textwidth]{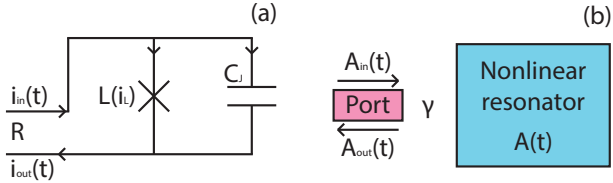}
\caption{(Color online) (a). Circuit diagram of the most basic JPA. (b). Description of JPA with the input-output formalism.}
\label{fig1}
\end{center}
\end{figure}

The most basic JPA model consists of a parallel lossless $LC$ resonant circuit where the Josephson junction plays a role of an inductor with a non-linear dependence on the current $L(i)$ and the capacitance $C$ comprises geometric capacitance of the circuit and the internal capacitance of the junction [Fig.~\ref{fig1}(a)]. Without loss of generality we consider that the resonant circuit is galvanically connected to a semi-infinite transmission line with characteristic impedance $R$ (note that in practice the impedance of the transmission line is typically fixed to 50 Ohm and the effective impedance seen by the $LC$ circuit is controlled by an additional coupling capacitor). 

With help of DC and AC Josephson effect equations, the Kirchhoff's circuit laws and the Telegrapher's equations theory of a transmission line one can obtain an equation for the phase of the Josephson junction as
\begin{equation}
\label{LC_Eq}
\dfrac{d^2\varphi}{d\tau^2}+\dfrac{1}{Q}\dfrac{d\varphi}{d\tau}+\sin\varphi = i_p \cos(\Omega\tau)+i_{s}(\tau),
\end{equation}
where the amplitude of the harmonic pump current $i_p$ and the signal current $i_s(\tau)$ are normalized to the critical current of the Josephson junction $I_c$, $Q = \omega_0 R C$ is the quality factor of the oscillator, $\omega_0=1/\sqrt{L_j C}$ is the natural resonance frequency of the circuit, $L_j = \Phi_0/(2\pi I_c)$ is the Josephson inductance and $\Phi_0=h/2e$ is the flux quantum. We also use dimensionless frequency $\Omega=\omega_p/\omega_0$ and time $\tau=\omega_0t$. By approximating  $\sin \varphi \simeq \varphi - \varphi^3/3$ one obtains the equation of motion for Duffing oscillator used extensively to analyse parametric amplifier properties \cite{Rhoads}.

After a standard quantization procedure the quantum  Hamiltonian of a capacitively shunted Josephson junction reads
\begin{equation}
\label{H}
\hat H = \frac{\hat q^2}{2 C} + E_j(1 - \cos \hat\varphi),
\end{equation}
where $\hat q$ is the operator of a charge on the capacitor and $E_j = I_c \Phi_0/(2\pi)$ is the Josephson energy.

By introducing creation and annihilation operators
\begin{equation}
\begin{array}{l}
\label{OpC_A}
\hat A  = \sqrt{\dfrac{C\omega_0}{2\hbar}}\left(\dfrac{\Phi_0}{2\pi}\hat\varphi +\dfrac{i}{C\omega_0}\hat q\right),\\
\hat A^{\dagger}  = \sqrt{\dfrac{C\omega_0}{2\hbar}}\left(\dfrac{\Phi_0}{2\pi}\hat\varphi-\dfrac{i}{C\omega_0}\hat q\right),
\end{array}
\end{equation}
and dropping the constant zero-point motion term we convert (\ref{H}) to 
\begin{equation}
\label{Exact_Hamiltonian}
\hat{H} = \hbar\omega_0\left(\hat A^{\dagger}\hat A+\sum_{k=2}^{\infty}\dfrac{2^{2k-3}}{(2k)!}\left(\dfrac{K}{\omega_0}\right)^{k-1}\left(\hat A+\hat A^{\dagger}\right)^{2k} \right).
\end{equation}
Here we  explicitly expressed the non-linearity as a sum of different orders of the weak effective Kerr non-linearity constant $K$ defined as
\begin{equation}
\label{Kerr_Constant}
K=-\dfrac{\omega_0}{8}\dfrac{2e\omega_0}{I_c}.
\end{equation}

Parametric amplification can be naturally studied in the framework of the input-output formalism \cite{Gardiner}, \cite{Walls}. By introducing incoming $\hat A_{in}$ and outgoing $\hat A_{out}$ microwave fields [Fig.~\ref{fig1}(b)] the intra-oscillator $\hat A$ field is governed by the following quantum stochastic differential equation
\begin{equation}
\label{IOT_Eq}
\dfrac{d\hat A}{dt}=-\dfrac{i}{\hbar}\left[\hat A(t),\hat{H}_{sys}\right]-\dfrac{\gamma}{2}\hat A(t)+\sqrt{\gamma}\hat A_{in}(t),
\end{equation}
where the coupling rate is given by $\gamma=\omega_0/Q$ and we neglected the internal losses of the $LC$ oscillator. The boundary conditions for all the fields at the input port of an oscillator read
\begin{equation}
\label{IOT_BC}
\hat A_{out}(t)=\sqrt{\gamma}\hat A(t)-\hat A_{in}(t).
\end{equation}
Substituting the Hamiltonian \eqref{Exact_Hamiltonian} into Eq \eqref{IOT_Eq} and taking into account the expression for the commutator $\left[A,\left(\hat A+\hat A^{\dagger}\right)^{2k}\right]=2k\left(\hat A+\hat A^{\dagger}\right)^{2k-1}$ for all integers $k\geq1$ 
we obtain the following exact equation of motion
\begin{multline}
\label{Eq_of_Motion}
\dfrac{d\hat A}{dt} =-i\omega_0\displaystyle\sum_{k=1}^{\infty}\dfrac{2^{2k-1}}{(2k+1)!}\left(\dfrac{K}{\omega_0}\right)^{k} \left(\hat A+\hat A^{\dagger}\right)^{2k+1}\\
-\left(\dfrac{\gamma}{2}+i\omega_0\right)\hat A(t)+\sqrt{\gamma}\hat A_{in}(t). \qquad\qquad\quad
\end{multline}

\section{Classical response and steady-state solution for pump field}
\label{ch_C_R}
An input field of JPA consists of a strong harmonic pump field at the pump frequency $\omega_p$ and a weak signal to be amplified. Extracting the harmonic time-dependence at the pump frequency $\omega_p$ explicitly we respectively write the input, intra and output fields as
\begin{equation}
\begin{array}{l}
\label{In_Intra_Out_Fields}
\hat A_{in}(t)=\left(\alpha_{in}+\hat a_{in}(t) \right)e^{-i\omega_pt},\\
\hat A(t)=\left(\alpha+\hat a(t) \right)e^{-i\omega_pt},\\
\hat A_{out}(t)=\left(\alpha_{out}+\hat a_{out}(t) \right)e^{-i\omega_pt}.
\end{array}
\end{equation}
Here the first terms on the right hand sides of \eqref{In_Intra_Out_Fields} represent the strong pump fields treated as classical complex amplitudes ($\alpha_{in}$, $\alpha$, $\alpha_{out}$)  while the other terms correspond to weak quantum signal fields described by operators ($\hat a_{in}$, $\hat a$, $\hat a_{out}$).

Neglecting weak signal contributions in \eqref{In_Intra_Out_Fields} and substituting only  the classical pump fields into \eqref{Eq_of_Motion} we obtain the following equation 
\begin{multline}
\label{Steady_state_sol}
-i\omega_p\alpha=-i\omega_0\alpha\displaystyle\sum_{k=1}^{\infty}\dfrac{2^{2k-1}}{k!(k+1)!}\left(\dfrac{K|\alpha|^2}{\omega_0}\right)^{k}\\
-\left(\dfrac{\gamma}{2}+i\omega_0\right)\alpha+\sqrt{\gamma}\alpha_{in}, \qquad\qquad\quad
\end{multline}
where we have taken into account the terms at the pump frequency only, applying the following approximation
\begin{equation}
\label{RWA}
\left(\alpha e^{-i\omega_pt}+\alpha^{\ast}e^{i\omega_pt}\right)^{2k+1}\rightarrow \dfrac{(2k+1)!}{k!(k+1)!}\alpha|\alpha|^{2k}e^{-i\omega_pt}.
\end{equation}
We note that the approximation \eqref{RWA} is equivalent to accounting only for terms with equal number of creation and annihilation operators in the Hamiltonian \eqref{Exact_Hamiltonian}. Eq.~\eqref{Steady_state_sol} has a harmonic steady-state solution at pump frequency $\omega_p$ and is a good approximation of an exact solution for {\it weak} non-linear processes.

It is well-known that the behaviour of the non-linear response (even for weak non-linearities) depends strongly on the amplitude of input field $\alpha_{in}$~\cite{Landau}, \cite{Jordan}. Specifically, the resonance frequency deviates from the natural frequency $\omega_0$ and decreases monotonically as the input field increases. At some critical value of the input pump field $\alpha_{in}^{crit}$ the response becomes bistable in some range of frequencies while below $\alpha_{in}^{crit}$ the fields have unique and stable solutions at any frequency. The bistable regime (above critical pump) is used in Josephson bifurcation amplifiers \cite{Vijay2} and is not considered in this manuscript. 

One may obtain the analytic expression for the critical value of an input field $\alpha_{in}^{crit}=\sqrt{-\gamma^2/\sqrt{27}K}$ (see for example \cite{Eichler1}) if only the leading (quartic) non-linear term in the Hamiltonian \eqref{Exact_Hamiltonian} is kept. It is convenient to normalize the input field to this value as $\alpha_{in}=r e^{i\phi}\alpha_{in}^{crit}$, where $r$ is the real scaling factor and $\phi$ is the initial phase of the input pump field. Multiplying both sides of equation \eqref{Steady_state_sol} by their complex conjugate parts we obtain the equation 
\begin{equation}
\begin{array}{l}
\label{Eq_n}
\left[\dfrac{1}{4Q^2}+\left(1-\Omega+\displaystyle\sum_{k=1}^{\infty}\dfrac{2^{2k-1}(-n)^k}{k!(k+1)!}\right)^2\right]n=\dfrac{r^2}{\sqrt{27}Q^3},
\end{array}
\end{equation}
where $\Omega=\omega_p/\omega_0$ is the dimensionless pump frequency and $n = -|\alpha|^2K/\omega_0$ is the normalised number of pump photons in the oscillator. Eq. \eqref{Eq_n} has a convenient form if one needs to keep only a certain number of the non-linear orders in the series. For example, keeping only the terms for $k = 1$ we obtain the well-known result for the cubic non-linearity.
One can use the following closed form of \eqref{Eq_n} to account for a complete $sine$ non-linearity
\begin{equation}
\label{Eq_n_Bess}
\left[\dfrac{1}{4Q^2}+\left(\dfrac{1}{2}-\Omega+\dfrac{J_1(4\sqrt{n})}{4\sqrt{n}}\right)^2\right]n=\dfrac{r^2}{\sqrt{27}Q^3},
\end{equation}
where $J_m(\cdot)$ is the Bessel function of the first kind of order $m$.

\begin{figure}[htbp]
\begin{center}
\includegraphics[width=0.5\textwidth]{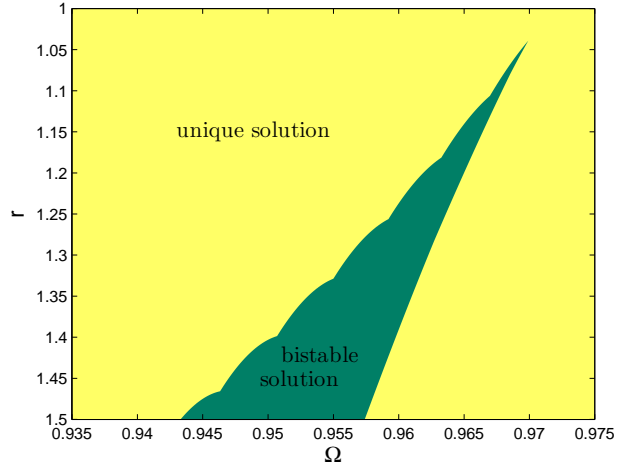}
\caption{(Color online) Frequency-amplitude ($\Omega,r$) stability diagram based on numerical solution of Eq.~\eqref{Eq_n_Bess} for a fixed value of the quality factor $Q=30$. Yellow region corresponds to a unique stable solution while the green region shows parameter range for three different solutions  (two stable and one unstable solutions). The most right point of the green domain $(0.96988,1.0401)$ is the bifurcation point for the complete $sine$ non-linearity.}
\label{SD}
\end{center}
\end{figure}

As mentioned above Eqs.~\eqref{Eq_n}-\eqref{Eq_n_Bess} are normalized to the critical input field corresponding to the leading non-linear term. In other words, the case $r=1$ corresponds to the bifurcation point for the cubic non-linearity only while the complete $sine$ non-linearity has the bifurcation point at some other value of the parameter $r$. In order to find this value one can plot a stability diagram [Fig.~\ref{SD}] revealing parameter domains corresponding to a unique solution and to the bistable mode of Eq.~\eqref{Eq_n_Bess}. 

Having found the number of photons $n$ we can express the amplitude of pump field $\alpha$ using equation \eqref{Steady_state_sol} as
\begin{multline}
\label{Alpha}
\alpha =\dfrac{1}{\sqrt{Q\omega_0}}\dfrac{re^{i\phi}\alpha_{in}^{crit}}{\dfrac{1}{2Q}+i\left(1-\Omega+\displaystyle\sum_{k=1}^{\infty}\dfrac{2^{2k-1}(-n)^k}{k!(k+1)!}\right)}\\
=\dfrac{1}{\sqrt{Q\omega_0}}\dfrac{re^{i\phi}\alpha_{in}^{crit}}{\dfrac{1}{2Q}+i\left(\dfrac{1}{2}-\Omega+\dfrac{J_1(4\sqrt{n})}{4\sqrt{n}}\right)}. \qquad\quad
\end{multline}

In experiment the JPA can be characterized  via measuring the reflection coefficient $S_{11}=\alpha_{out}/\alpha_{in}$. Using the boundary conditions \eqref{IOT_BC} and equation \eqref{Alpha} we obtain
\begin{equation}
\label{S11}
S_{11} = \dfrac{\dfrac{1}{2}-iQ\left(\dfrac{1}{2}-\Omega+\dfrac{J_1(4\sqrt{n})}{4\sqrt{n}}\right)}{\dfrac{1}{2}+iQ\left(\dfrac{1}{2}-\Omega+\dfrac{J_1(4\sqrt{n})}{4\sqrt{n}}\right)}.
\end{equation}

\section{Theory of linear amplification}
\label{ch_T_L}
As the powers of the input signal $\langle a^{\dagger}_{in}a_{in}\rangle$ and amplified signal $\langle a^{\dagger}_{out}a_{out}\rangle$ are typically much smaller than the one of the input pump field $|\alpha_{in}|^2$, presence of a weak quantum signal does not affect classical solution \eqref{Alpha}. Known as {\it the stiff pump approximation} \cite{Kamal}, this assumption allows us to derive the equation of motion for the signal in the linear response regime. 

One can substitute Eqs.~\eqref{In_Intra_Out_Fields} into Eq.~\eqref{Eq_of_Motion} and keep only the terms linear in quantum signal $a$, $a^{\dagger}$. Applying both approximation \eqref{RWA} for the pump signal and linearisation procedure for the quantum signal one can get
\begin{multline}
\label{Lin_Procedure}
\left(\hat A+\hat A^{\dagger}\right)^{2k+1}=\left(\alpha e^{-i\omega_pt}+\alpha^{\ast}e^{i\omega_pt}\right)^{2k+1}\\
+(2k+1)\left(\alpha e^{-i\omega_pt}+\alpha^{\ast}e^{i\omega_pt}\right)^{2k} (\hat a(t)e^{-i\omega_pt}\\
+\hat a^{\dagger}(t)e^{i\omega_pt})+O(\hat a^2)\longrightarrow (2k+1)!\left(\dfrac{\alpha|\alpha|^{2k}}{k!(k+1)!}\right.\\
+\left. \dfrac{|\alpha|^{2k}}{(k!)^2}\hat a(t)+\dfrac{\alpha^2|\alpha|^{2(k-1)}}{(k-1)!(k+1)!}\hat a^{\dagger}(t)\right)e^{-i\omega_pt}.
\end{multline}
Using Eq.~\eqref{Lin_Procedure} we obtain the following linearised equation of motion for the signal
\begin{equation}
\label{Lin_Eq_of_Mot}
\dfrac{d\hat a}{d\tau}=l_1\hat a(\tau)+l_2\hat a^{\dagger}(\tau)+\dfrac{\hat a_{in}(\tau)}{\sqrt{Q\omega_0}},
\end{equation}
where the coefficients are defined as follows
\begin{eqnarray}
\label{l1}
l_1 &=& i\left(\Omega-\dfrac{1}{2}-\dfrac{J_0(4\sqrt{n})}{2}\right)-\dfrac{1}{2Q},
\\\label{l2}
l_2 &=& -i\frac{\alpha^2 K}{\omega_0}\dfrac{J_2(4\sqrt{n})}{2n}.
\end{eqnarray}

Eq.~\eqref{Lin_Eq_of_Mot} together with its Hermitian conjugate equation forms a system of two linear differential equations with time-independent coefficients and can be solved in the frequency domain. Without loss of generality we consider the amplification process of a signal at some frequency $\omega_s$ and write the time dependence of intra resonator field as superposition of their Fourier components
\begin{equation}
\label{Four_Comp}
\hat a(\tau)=\dfrac{1}{\sqrt{2\pi}}\int_{-\infty}^{\infty}\hat a(\Delta)e^{-i\Delta\tau}d\Delta,
\end{equation}
\begin{equation}
\label{Four_Comp_2}
\hat a^{\dagger}(\tau)=\dfrac{1}{\sqrt{2\pi}}\int_{-\infty}^{\infty}\hat a^{\dagger}(-\Delta)e^{-i\Delta\tau}d\Delta,
\end{equation}
where $\Delta=(\omega_s-\omega_p)/\omega_0$ is the normalized detuning of the signal from the pump frequency. Applying Fourier transform to Eq.~\eqref{Lin_Eq_of_Mot} and its Hermitian conjugate pair we eventually arrive at the system of two linear algebraic equations with the  solution
\begin{equation}
\label{Lin_Sol}
\begin{array}{l}
\hat a(\Delta)=\dfrac{1}{\sqrt{Q\omega_0}}\dfrac{-(l_1^{\ast}+i\Delta)\hat a_{in}(\Delta)+l_2\hat a_{in}^{\dagger}(-\Delta)}{(l_1+i\Delta)(l_1^{\ast}+i\Delta)-l_2l_2^{\ast}},\\
\\
\hat a^{\dagger}(-\Delta)=\dfrac{1}{\sqrt{Q\omega_0}}\dfrac{l_2^{\ast}\hat a_{in}(\Delta)-(l_1+i\Delta)\hat a_{in}^{\dagger}(-\Delta)}{(l_1+i\Delta)(l_1^{\ast}+i\Delta)-l_2l_2^{\ast}}.
\end{array}
\end{equation}
Applying the boundary conditions \eqref{IOT_BC} in the frequency domain we express the output signal Fourier component via the input one as
\begin{equation}
\label{IOT_BC_FD}
\hat a_{out}(\Delta)=g(\Delta)\hat a_{in}(\Delta)+m(\Delta)\hat a^{\dagger}_{in}(-\Delta),
\end{equation}
where we defined
\begin{equation}
\label{Lin_Gain}
\begin{array}{l}
g(\Delta)=-\dfrac{1}{Q}\dfrac{l_1^{\ast}+i\Delta}{(l_1+i\Delta)(l_1^{\ast}+i\Delta)-l_2l_2^{\ast}}-1,\\
\\
m(\Delta)=\dfrac{1}{Q}\dfrac{l_2}{(l_1+i\Delta)(l_1^{\ast}+i\Delta)-l_2l_2^{\ast}}.
\end{array}
\end{equation}
Coefficients $g(\Delta)$ and $m(\Delta)$ satisfy the relation $G(\Delta)=|g(\Delta)|^2=|m(\Delta)|^2+1$ where $G(\Delta)$ can be identified as the gain of phase-insensitive linear amplifier~\cite{Eichler1}.

In the traditional JPA theory the gain is calculated in cubic approximation by keeping only the first term $k = 1 $ in Eqs.~\eqref{Eq_n} and \eqref{Alpha}. Now we have the exact expressions for the gain \eqref{Lin_Gain} where we account for all non-linear terms in the equation of motion \eqref{Eq_of_Motion}.
Using this result we can analyse the effect of higher order non-linearities on the gain of JPA.
 
\begin{figure}[htbp]
\begin{center}
\includegraphics[width=0.5\textwidth]{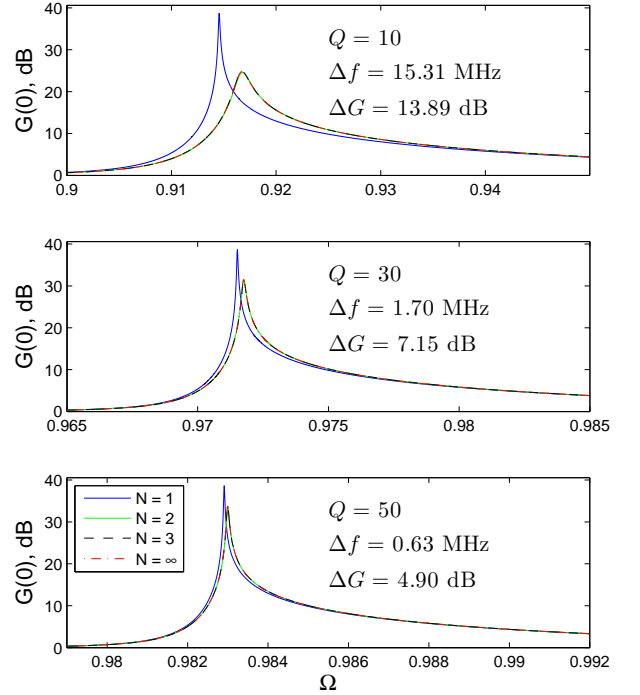}
\caption{(Color online) Gain as function of pump frequency for the fixed pump power $r=0.99$, zero signal detuning $\Delta=0$ and different quality factors $Q$ of an oscillator.  We also choose $f_0=7$~GHz and critical current $I_c=2~\mu A$. The colour represents different number of terms $N$ taken into account in equation \eqref{Eq_of_Motion}. $N = 1$ corresponds to the known case of qubic non-linearity, $N = 2, 3$ account for the next two higher quintic and septic non-linearities, $N = \infty$ represents the solution for a complete $\sin \phi$ non-linearity. $\Delta f$ indicates the frequently difference between the maximum of gain and $\Delta G$ shows the difference in maximal attainable gain for $N = 1$ and one for the exact solution $N = \infty$.} 
\label{fig2}
\end{center}
\end{figure}

In the Fig.~\ref{fig2} the gain of JPA at zero detuning $G(0)$ is plotted as a function of the dimensionless pump frequency $\Omega$ for the fixed pump power close to the critical value and different values of the $Q$-factor. For each value of $Q$-factor we take into account different number of non-linear terms $N$. To make a link to a real experimental situation we choose typical values for the natural frequency $f_0=\omega_0/2\pi=7$~GHz and the critical current of the Josephson junction $I_c=2~\mu A$ and adjust the pump power to recover the experimentally relevant values of maximal gains. 

Our results show that higher orders of non-linearity lead to a decrease and a frequency shift of the maximal achievable gain for a given pump strength and a quality factor. As the quality factor decreases the steady state normalized photon number $n$ corresponding to the critical pumping strength increases (see Eq.~\eqref{Eq_n_Bess}) and one needs to take into account more terms in the expansion of the Bessel function (in physical terms it means that the current through the Josephson junction start to approach the critical current $I_c$). That explains our observation that the effect is more pronounced for lower values of $Q$ and is practically negligible for $Q > 150$ (not shown). 
Within relevant parameter range the main contribution to the difference between the known solution for the fundamental cubic non-linearity $(N=1)$ and the exact solution is provided by the next higher quintic $(N=2)$ non-linearity while the contributions from septic $(N=3)$ and all other orders  are not substantial.

\begin{figure}[htbp]
\begin{center}
\includegraphics[width=0.5\textwidth]{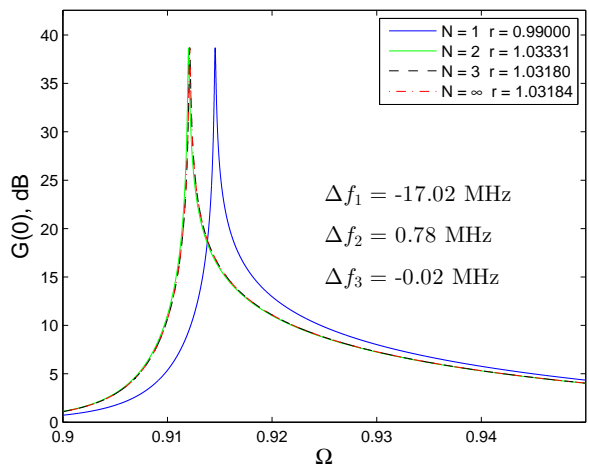}
\caption{(Color online) Signal gain dependences on pump frequency for the fixed $Q=10$, zero detuning $\Delta=0$, natural frequency $f_0=7$~GHz and critical current $I_c=2~\mu A$. The colour represents different number of terms $N$ taken into account in equation of \eqref{Eq_of_Motion}. The pump amplitude $r$ was adjusted to achieve the same values of maximal gain.  $\Delta f_n$ with $n = 1,2,3$ indicate corresponding frequency shifts of the maximal gain between solutions for $N = \infty -1$; $N = \infty -2$ and $N = \infty -3$, respectively.}
\label{fig3}
\end{center}
\end{figure}

For any value of $Q$-factor we can reduce the difference between maximal values of the signal gain by varying the level of applied pump power. In  Fig.~\ref{fig3} one can see the gain dependences on pump frequency calculated for the fixed $Q=10$, zero detuning $\Delta  =0$ and different number of non-linear terms $N$. For each of the solutions the pump amplitude was adjusted to achieve the same value of maximal gain while all other parameters were kept constant. Our results show that the shape of the curves are virtually identical and the only effect of the non-linear terms of higher orders is a shift of the resonance frequency (gain profile) of the JPA. 

In experiment it is hard to calibrate the exact value of power applied to a device in the absolute scale. It is also difficult to characterize the frequency shift due to the high-order non-linearities as the resonance frequency of an oscillator depends on the pumping power. This suggests that the contribution of higher-order non-linearities to gain of JPA in the linear amplification regime is not very significant and hard to observe experimentally.

\section{Gain saturation effect}
\label{ch_G_S}
The linearised signal and idler gain expressions \eqref{Lin_Gain} are valid as long as the output power of amplified fields is much smaller than the pump field power. Firstly, we estimate gain saturation effect due to the fundamental cubic non-linearity. Using Eq.~\eqref{Eq_of_Motion} with only one term in the sum for $k = 1$ and Eq.~\eqref{In_Intra_Out_Fields}, where the classical pump field $\alpha$ is already found via Eqs.~\eqref{Eq_n_Bess}-\eqref{Alpha}, we obtain the following equation of motion for quantum signal in the cubic approximation
\begin{multline}
\label{Nonlin_Eq_of_Mot}
\dfrac{d\hat a}{d\tau}=l_1\hat a+l_2\hat a^{\dagger}+c_1\hat a^2+c_2(\hat a\hat a^{\dagger}+\hat a^{\dagger}\hat a)\\
+c_3(\hat a^2\hat a^{\dagger}+\hat a\hat a^{\dagger}\hat a+\hat a^{\dagger}\hat a^2)+\dfrac{\hat a_{in}(\tau)}{\sqrt{Q\omega_0}}. \qquad\qquad
\end{multline}
The linear coefficients $l_1$ and $l_2$ are given in \eqref{l1} and \eqref{l2} while the cubic coefficients are defined as
\begin{eqnarray}
\label{c1}
c_1&=&-i\frac{K}{\omega_0}\alpha^{\ast}\dfrac{J_1(4\sqrt{n})}{2\sqrt{n}},\\
\label{c2}
c_2&=&\frac{\alpha}{\alpha^*}c_1,\\
\label{c3}
c_3&=&-i\dfrac{K}{\omega_0}\dfrac{J_0(4\sqrt{n})}{3}.
\end{eqnarray}
We note that to derive Eq.~\eqref{Nonlin_Eq_of_Mot}, an approximation similar to \eqref{Lin_Procedure} is used again, where both the linear terms of quantum signal and the three non-linear terms corresponding to the cubic non-linearity are taken into account. In this way one can derive more precise equations for the quantum signal containing the non-linear terms of higher orders. Also, we can include all the non-linear terms in Eq.~\eqref{Lin_Eq_of_Mot} to relax the restriction on smallness amplitude of input signal. However, in this case the signal equation becomes cumbersome and not portable. Fortunately, we can simplify the signal equation conducting the derivation of the signal equation in the semi-classical limit, i.e. we take $a(t)$ to be a complex function and not an operator. Mathematically, it results in the two following approximations
\begin{multline}
\label{Nonlin_Procedure}
\left(\hat A+\hat A^{\dagger}\right)^{2k+1}\longrightarrow\left[\left(\alpha+a(t)\right)e^{-i\omega_pt}\right.\\
\qquad\left.+\left(\alpha^{\ast}+a^{\ast}(t)\right)e^{i\omega_pt}\right]^{2k+1}\\
\longrightarrow\dfrac{(2k+1)!}{k!(k+1)!}\left(\alpha+a(t)\right)^{k+1}\left(\alpha^{\ast}+a^{\ast}(t)\right)^ke^{-i\omega_pt}.
\end{multline}
On the first step we go to the semi-classical limit while on the second step we account for the terms ar the pump frequency only. Applying the approximations \eqref{Nonlin_Procedure} to each of the non-linear terms in Eq.~\eqref{Eq_of_Motion} we finally get the exact non-linear signal equation in the semi-classical limit as
\begin{multline}
\label{Nonlin_Eq_of_Mot_Full}
\dfrac{da}{d\tau}=-\left[\dfrac{1}{2Q}+i\left(\dfrac{1}{2}-\Omega+\dfrac{J_1(4\sqrt{m(a)})}{4\sqrt{m(a)}}\right)\right]a(\tau)\\
+i\left[\dfrac{J_1(4\sqrt{n})}{4\sqrt{n}}-\dfrac{J_1(4\sqrt{m(a)})}{4\sqrt{m(a)}}\right]\alpha+\dfrac{a_{in}(\tau)}{\sqrt{Q\omega_0}},
\end{multline}
where $m(a)=n-K(\alpha^{\ast}a+\alpha a^{\ast}+a^{\ast}a)/\omega_0$. 
 
We used the fourth-order finite-difference (Runge-Kutta) method to solve the system of Eq.~\eqref{Nonlin_Eq_of_Mot_Full} and its complex conjugate pair as well as the system of the approximate equation of motion \eqref{Nonlin_Eq_of_Mot} and its Hermitian conjugate pair in the semi-classical limit. To calculate the gain of JPA we choose the excitation on the right hand side of Eq.~\eqref{Nonlin_Eq_of_Mot_Full} [or Eq.~\eqref{Nonlin_Eq_of_Mot}] in the form of a harmonic signal at some frequency $\omega_s$: $a_{in}(\tau)=a_{in}(\Delta)e^{-i\Delta\tau}$ while in the complex conjugate pair of Eq.~\eqref{Nonlin_Eq_of_Mot_Full} [or \eqref{Nonlin_Eq_of_Mot}] we set $a^\dagger_{in}(-\Delta) = 0$. By numerically calculating $a_{out}$ at the same frequency $\omega_s$ we obtain $a_{out} (\Delta)$ and evaluate the gain using Eq.~\eqref{IOT_BC_FD}. 

\begin{figure}[htbp]
\begin{center}
\includegraphics[width=0.5\textwidth]{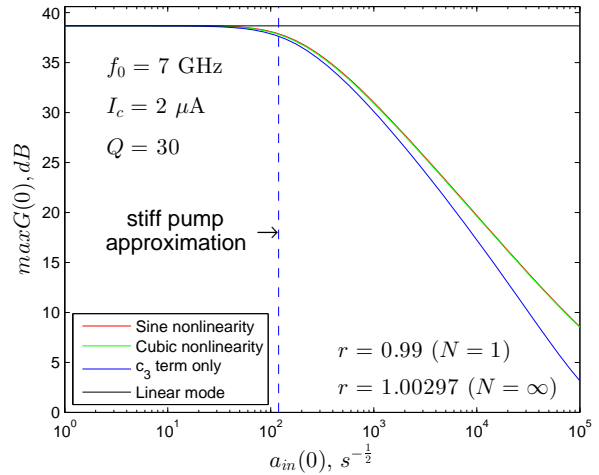}
\caption{(Color online) Signal gain saturation due to $sine$ and cubic non-linearities.
Here the natural frequency $f_0=7$~GHz, the critical current $I_c = 2\mu A$ and the quality factor $Q=30$ of the JPA are the same as in the linear case depicted in the middle panel of Fig.~\ref{fig2}.
}
\label{fig4}
\end{center}
\end{figure}

In Fig.~\ref{fig4} we present the  maximal signal gain $\max G(0)$ as function of the input signal amplitude $a_{in}(0)$ corresponding to the zero frequency detuning $\Delta=0$ between the signal $\omega_s$ and the pump $\omega_p$ frequencies. We use similar parameters of JPA as in the middle panel of Fig.~\ref{fig2}. The red curve represents the solution of exact equation of motion~\eqref{Nonlin_Eq_of_Mot_Full} while the  green curve correspond to  the solution of Eq.~\eqref{Nonlin_Eq_of_Mot} in the semi-classical approximation. We also solve Eq.~\eqref{Nonlin_Eq_of_Mot} with $c_3$ term only (blue curve in Fig.~\ref{fig4}) by setting manually $c_1=c_2=0$. For the solution of \eqref{Nonlin_Eq_of_Mot} we set the pump power close to the bifurcation point putting $r = 0.99$. In order to achieve the same gain level in the case when all non-linear terms ($N=\infty$) are taken into consideration in the signal equation \eqref{Nonlin_Eq_of_Mot_Full} we slightly change the pump power to $r = 1.00297$ (see discussion in the previous section). The analytical solution \eqref{Lin_Gain} corresponding to the gain in linear amplification regime is shown as a horizontal line for reference.

As expected, Fig.~\ref{fig4} demonstrates that the gain decreases at large powers of input signals. The vertical dotted line in Fig.~\ref{fig4} corresponds to the relation $P_{in}^{pump}/P_{out}^{signal}=20~dB$ which represents the practically found rule for validity of the stiff pump approximation. We can also conclude that the fundamental contribution comes from the third term in the cubic non-linearity $3c_3a^{\ast}a^2$ (blue line in Fig.~\ref{fig4}). 

\begin{figure}[htbp]
\begin{center}
\includegraphics[width=0.5\textwidth]{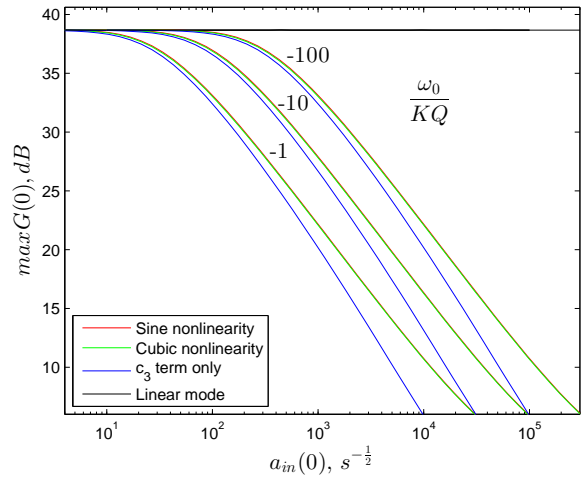}
\caption{(Color online) Dependence of the dynamic range on the Kerr non-linearity under constant the natural frequency $f_0=7$~GHz and the quality factor $Q=30$.}
\label{fig5}
\end{center}
\end{figure}

It is logical to expect that the dynamic range increases with decreasing the coefficients $c_1$, $c_2$, $c_3$. Using the formula \eqref{c3}, expressing the critical current as $I_c=\omega_0^2C_j\varphi_0$, taking into account \eqref{Kerr_Constant} and the quality factor formula $Q=\omega_0C_jR$ we find ${c_i}\propto K/\omega_0 \sim 1/Q$. In order to illustrate this dependence we plot the maximal gain for three different values of the ratio $\omega_0/KQ=-1$, $-10$, $-100$ [Fig.~\ref{fig5}] by adjusting the critical current values keeping other parameters constant. It is evident that the JPA dynamical range increases with increasing the ratio $\omega_0/|K|Q$. 

Our solution is based on approximation that both the input pump field and the intra resonator (output) pump field have the same harmonic dependence in time (see Eq.~(\ref{RWA})) which is justified for  $i_p=I_p/I_c << 1$. Indeed, equation \eqref{LC_Eq} always has a period-one (harmonic) steady state solution at the pump frequency if the condition of small oscillations $I_p/I_c<<1$ is satisfied \cite{Jordan}, \cite{Bogoliubov}. With increasing the ratio $I_p/I_c$ the form of solution to Eq.~(\eqref{LC_Eq}) can be changed dramatically demonstrating a period doubling cascade through a sequence of pitchfork bifurcations (transition from regular to chaotic dynamics) and eventually chaotic behaviour~\cite{Jordan}. Under these conditions the classical solution for intra resonator field (\ref{Alpha}) fails independently of the number of non-linear terms taken into account. 
The number of pump photons in the JPA resonator at the point close to bifurcation is proportional to $N_p\sim\alpha_{crit}^{2}\sim I_c/Q^2$. Therefore, the pump current approaches  the critical current when sufficiently low quality factor is chosen affecting the expected operation of the amplifier and imposing limitation on its dynamic range and bandwidth~\cite{Eichler1}.

\section{Conclusion}
\label{ch_C}
We obtained the expression for the gain of JPA including all orders of non-linearity in the linear amplification regime.
Comparing to the conventional solution for the qubic non-linearity we found that higher order terms lead to a decrease of the maximum achievable gain for given pump power and to a frequency shift of maximum gain point. By increasing the pump power one can restore the value of the maximal gain and only the frequency shift of the maximal gain point remains. 

To study the effect of the gain saturation at high signal powers we solved the equation of motion including all non-linearities  in the semi-classical approximation beyond linear response regime.
We revealed the principal term responsible for the gain saturation effect and identified the parameter controlling the dynamic range of JPA.

Our solution is valid when both the input pump fields inside and outside of the resonator have the same harmonic dependence which is the case when the current though the Josephson junction is much less than the critical current  $I_p/I_c << 1$.  In this case of weak non-linear process the equation of motion always has a period-one (harmonic) steady-state solution at the pump frequency and our approximation is justified.

At higher driving conditions the solution to Eq \eqref{LC_Eq} may demonstrate a period doubling cascade through a sequence of pitchfork bifurcations (like it takes place for the Duffing equation  \cite{Jordan}) and our approach (as well as the conventional cubic formula) will be inaccurate. It is interesting to note that at higher driving conditions the solution to the Duffing equation has a parameter domain when a large amplitude period-one steady state solution \cite{Jordan}. Therefore, Eq.~\eqref{LC_Eq} can in principle also have the same properties. We speculate that in this case conditions for JPA operation can be found even for $I_p\lesssim I_c$ which will allow simultaneous large bandwidth and large dynamic range. This will be a subject of future investigation.

\begin{acknowledgements}
We would like to express our appreciation and thankfulness to Dr.~C.~Eichler and Dr.~N.~Pakhira for their valuable comments and suggestions for improvements of the paper.

This work was supported by the Australian Research Council (Grant No. CE110001013).
\end{acknowledgements}


\end{document}